\documentclass[showpacs,amsmath,amssymb,prl,superscriptaddress,footinbib,twocolumn]{revtex4-1}
\usepackage{amssymb}
\usepackage{graphics}
\usepackage{epstopdf}
\usepackage{epsfig}
\usepackage{dcolumn}
\usepackage{bm}
\renewcommand{\vec}[1]{\mathbf{#1}}
\newcommand{\abs}[1]{\left| #1 \right|} % for absolute value
\let\baraccent=\= % rename builtin command \= to \baraccent
\renewcommand{\=}[1]{\stackrel{#1}{=}} % for putting numbers above =
\newcommand{\pddc}[3]{\frac{\partial^2 #1}{\partial #2 \partial #3}} 
\usepackage{color}
 \definecolor{blue}{rgb}{0,0,1} %%{\color{blue} text...}
 \definecolor{sepia}{rgb}{0,0.8,0.2}
 \definecolor{redi}{rgb}{0.5176,0.0078,0.0078}

\begin{document}

\title{Degeneracy and Inversion of Band Structure for Wigner Crystals on a Toroidal Helix}

%\affiliation{Department of Physics, University of Athens, GR-15874 Athens, Greece}

%\email[]{azampeta@physnet.uni-hamburg.de}

\author{A. V.  Zampetaki}
\author{J. Stockhofe}
\affiliation{Zentrum f\"{u}r Optische Quantentechnologien, Universit\"{a}t Hamburg, Luruper Chaussee 149, 22761 Hamburg, Germany}
\author{P. Schmelcher}
\affiliation{Zentrum f\"{u}r Optische Quantentechnologien, Universit\"{a}t Hamburg, Luruper Chaussee 149, 22761 Hamburg, Germany}
\affiliation{The Hamburg Centre for Ultrafast Imaging, Luruper Chaussee 149, 22761 Hamburg, Germany}

\date{\today}

\begin{abstract}
We explore the formation of Wigner crystals for charged particles on a toroidal helix. Focusing on certain commensurate cases we show that the ground state undergoes
a pitchfork bifurcation from the totally symmetric polygonic to a zig-zag-like configuration with increasing radius of the helix.  
Remarkably, we find that for 
a specific value of the helix radius, below the bifurcation point, the vibrational frequency spectrum collapses to a single frequency.
This allows for an essentially independent small-amplitude motion of the individual particles and consequently localized excitations 
can propagate in time without significant spreading. Increasing the radius beyond the degeneracy point, the band structure is inverted,
with the out-of-phase oscillation mode becoming lower in frequency than the mode corresponding to the center of mass motion.

\end{abstract}

\maketitle

\textit{Introduction} At low temperatures and densities trapped charged particles tend to arrange in the so-called Wigner crystal \cite{Wigner1934}.
Such crystalline structures are studied extensively due to their applications in spectroscopy \cite{Thomson1993,Hermanspahn2000} and 
their connection with realizations of quantum simulators \cite{Johanning2009, Blatt2012} and quantum information processors \cite{Cirac1995, Kaler2003}.
When the trapping potential is harmonic, as approximately true for Penning \cite{Penning1936} or Paul traps \cite{Paul1990},
successive structural phase transitions occur while increasing the dimensionality of the system controlled by the transverse trapping frequency.
In particular, the system undergoes a transition from a linear string configuration to a planar zig-zag and finally to a three dimensional (3D)
helical structure \cite{Hasse1990,Birkl1992}. A type of linear to zig-zag transition has also been identified for ions trapped in octupole traps
\cite{Yurtsever2011,Cartarius2013}. The features of such transitions have been explored in detail
\cite{Cartarius2013,Fishman2008,Shimshoni2011,Mehta2013} since they constitute prototypical examples for the behaviour of condensed matter systems.

On the other hand, recent advances in nanofabrication have allowed  the construction of nanotubes with curved geometries such as rolls, spirals and helices 
 \cite{Prinz2000,Schmidt2001}. Helical traps have also been developed experimentally for ultra-cold neutral
atoms using counter-propagating Laguerre-Gaussian beams \cite{Bhatt2007, Okulov2012} or the evanescent field of a nanofiber \cite{Vetsch2010,Reitz2012}.
Studying the helical geometry is of fundamental importance since it constitutes a principal feature of structures
commonly appearing in nature, with prominent examples being amino-acids and the DNA molecule. Constraining interacting particles
to such a geometry gives rise to intriguing effects. For dipolar particles a peculiar quantum phase transition from liquid to gas \cite{Law2008} and the formation of 
crystalline chains \cite{Pedersen2014} are predicted. Identical charged particles on a helix are exposed to an effective oscillatory force leading to
 multiple classical bound states \cite{Kibis1992, Schmelcher2011, Zampetaki2013} despite the purely repulsive interaction in three dimensions. %Still, related studies of many-body systems are rare.

Here we investigate the structure and dynamics of Wigner crystals confined on a toroidal helix.
Such a system, owing to its intricate geometry, has the peculiar feature of a `mixed dimensionality': 
The classical particle motion is constrained to the 1D confining manifold whereas interactions take place through the 3D surrounding space \cite{Schmelcher2011}. 
We find that tuning the geometry induces a zig-zag-like transition, although the single particle configuration space remains strictly 1D.
Due to the 1D constraint that restricts the allowed excitations, this transition is accompanied by an unconventional deformation of the corresponding dispersion relation: For finite systems, 
there is a regime of inverted dispersion, with the out-of-phase mode being lowest in frequency.
Even more surprisingly, the transition passes through a stage where the complete linearization spectrum is essentially degenerate, such that \emph{any} low amplitude mode is an eigenmode of the system.
Notably, localized excitations do not transfer energy into the rest of the chain.

\begin{figure}[htbp]
\begin{center}
\includegraphics[width=6.8cm]{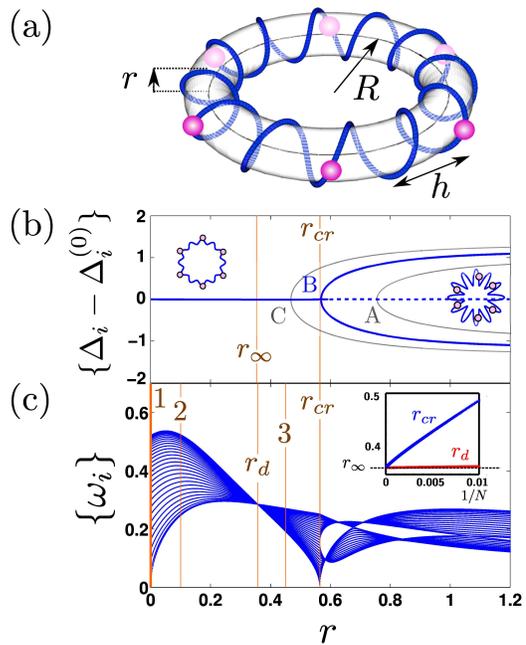}
\end{center}
\caption{\label{tore1} (color online). (a) Equidistant configuration of ions confined on the toroidal helix for $\nu=\frac{1}{2}$ and $N=6$.
(b) Equilibrium displacements of particles in stationary configurations as a function of the helix radius $r$ for filling $\nu=\frac{1}{2}$ and different numbers of particles: (A) $N=30$, (B) $N=60$, (C) $N=120$.
All values of $\Delta_i - \Delta_i^{(0)}$ are plotted on top of each other: For the polygonic configuration, all nearest-neighbor interparticle distances are identical, $\Delta_i = \Delta_i^{(0)}$ for all $i$,
while in the zig-zag-like configuration precisely two different distances are found, see the insets.
 The vertical lines indicate $r_\text{cr}$ for $N=60$ and the limiting value $r_\infty$ of $r_\text{cr}$ in the thermodynamic limit.
(c) Linearization spectrum as a function of $r$ of the stable solution (polygonic for $r<r_\text{cr}$, zig-zag-like for $r>r_\text{cr}$) for $N=60$. Degeneracy of all vibrational modes is observed at $r_\text{d}$.
The inset depicts how $r_\text{cr}$ and $r_\text{d}$ converge to a common value $r_\infty$ in the thermodynamic limit. }

\end{figure}

\textit{Toroidal helix} We consider a system of $N$ identical charged particles interacting via the repulsive Coulomb interaction and confined 
to move on a 1D toroidal helix, parametrized as
\begin{equation}
 \vec{r} (u)= \begin{pmatrix}
 \left( R+r \cos(u)\right)\cos(au) \\
 \left( R+r \cos(u)\right)\sin(au)\\
 r\sin(u)
\end{pmatrix}, \quad u \in [0, 2 M \pi],
\label{te1}
\end{equation}
with $R$ the major radius of the torus, $r$ the radius of the helix (minor radius of the torus)
and $h$ the helix pitch. The parameter $a=\frac{1}{M}$ is the inverse number of windings $M=\frac{2 \pi R}{h}$ (see Fig.\ref{tore1} (a)).
The effective Coulomb potential resulting from the confinement is given by 
$V(u_1,u_2,\ldots u_N)=\frac{1}{2}\sum_{i,j=1,i \neq j}^{N}\frac{\lambda}{\abs{\vec{r}(u_i)-\vec{r}(u_j)}}$. %\label{eq:coulg}

We perform a scaling transformation \cite{Zampetaki2013} resulting in the coupling constant $\lambda$ and the particle mass being set to $1$ in the following, while the helix pitch is fixed at $h=\pi/2$.
Note that the center of mass (CM) degree of freedom is coupled to the relative coordinates $\Delta_i=u_{i+1}-u_i$ for $r \neq 0$ since the confining manifold is not
a homogeneous helix \cite{Zampetaki2013}. For $r=0$, on the other hand, one recovers the confinement on a ring with radius $R$ 
where CM separation holds. In this ring limit, there is a single stable ground
state, namely the totally symmetric polygonic configuration. In contrast, the potential landscape of charged particles confined on helical manifolds
is very complex allowing for a large number of stable states for given parameters \cite{Schmelcher2011,Zampetaki2013}.

Here, we focus on the low energy crystalline configurations and their equilibrium properties for
systems with an even number of particles $N$ which divides the number of windings $M$,
i.e. $M=nN$ with $n=1,2,\ldots$ and $\nu=1/n \leq 1$ the filling factor. Then, the polygonic configuration 
$u_j^{(0)}=2\left(j-1\right) \pi n$ of the ring persists as a (stable or unstable) equilibrium configuration for all values of  $r$ 
with the charges being located equidistantly, $\Delta_j^{(0)}=2 \pi n$, at the outer circle of the toroidal helix (Fig.\ref{tore1} (a)).
However, for sufficiently large $N$  (e.g. $N>4$ for $\nu =\frac{1}{2}$) this configuration loses its stability at a finite $r=r_\text{cr}$ and undergoes a pitchfork bifurcation leading through symmetry breaking to a zig-zag-like
configuration (Fig.\ref{tore1} (b)) in which successive particles move in pairs to positive and negative values of the $z$-coordinate of the vector $\vec r$ (Eq. (\ref{te1})). 
For a fixed filling factor, 
here $\nu=\frac{1}{2}$, the bifurcation point $r_\text{cr}$ shifts to lower values of $r$ with increasing $N$ (thus also increasing $M$), tending 
to a finite value $r_{\infty}$ (Fig. \ref{tore1} (c) (inset)) in this thermodynamic limit,  with a rather slow convergence rate.
Surprisingly,
it turns out that the value of $r_{\infty}$ depends only on the pitch of the helix $h$, namely  $r_\infty=\frac{h}{\sqrt{2} \pi}$ in physical units (or $r_\infty = \frac{1}{2 \sqrt{2}}$ in our dimensionless units), independently of $\nu$.

Following the stable branch of solutions, we calculate the spectrum of vibrational modes in the harmonic approximation (Fig.\ref{tore1} (c)).
Intriguingly, in the regime $r< r_\text{cr}$ (where the stable configuration is still symmetric) this spectrum exhibits a crossing point $r_\text{d}$
where all modes are very close to degenerate.
The value $r_\text{d}$ also depends on the size of the system, decreasing for large $N$ and 
tending to $r_\infty$ in the thermodynamic limit (Fig. \ref{tore1} (c) (inset)), but much faster than $r_\text{cr}$ does. 
Thus, for finite systems an interval  $r_\text{d}<r<r_\text{cr}$ always exists.
In this region the spectrum is inverted, and finally the lowest eigenvalue crosses zero at $r_\text{cr}$, rendering the symmetric configuration unstable and leading to the observed pitchfork bifurcation (Fig. \ref{tore1} (b)). 
For $r>r_\text{cr}$ two branches of frequencies separated by a gap are created, as a result of the new emergent solutions possessing a doubled unit cell, whose deformation 
continues with increasing $r$.

\textit{Vibrational analysis} Let us now return to the frequency spectrum of the symmetric, polygonic configuration for $r<r_d$. 
This being a Wigner crystal with a one-particle unit cell, the corresponding dispersion relation consists of a single branch.
For its evaluation, we introduce the arc length parametrization
in which the kinetic energy and Euler-Lagrange equations assume the standard form \cite{Zampetaki2013}.
The dispersion relation then reads
\begin{equation}
 \omega^2(k)=\frac{1}{a^2((R+r)^2+r^2)}\sum_{l=1}^{N} H_{1,l} \exp \left(\frac{-ik(l-1){\Delta} s}{N}\right) \label{di1},
\end{equation}
with the Hessian at the equilibrium configuration $H_{i,j} = \pddc{V}{u_{i}}{u_{j}}\big|_{\{u_j^{(0)}\}}$ (we can fix one of its indices for symmetry reasons).
The prefactor in Eq. (\ref{di1}) results from transforming to the arc length $s$, $\Delta s$ denotes the arc length inter-particle distance of the symmetric solution and $k= \frac{2 \pi m}{N \Delta s}, ~ (m=0,\pm 1,\ldots \pm \frac{N}{2})$ 
is the wave number of the corresponding excitation.

\begin{figure}[htbp]
\begin{center}
\includegraphics[width=8.6cm]{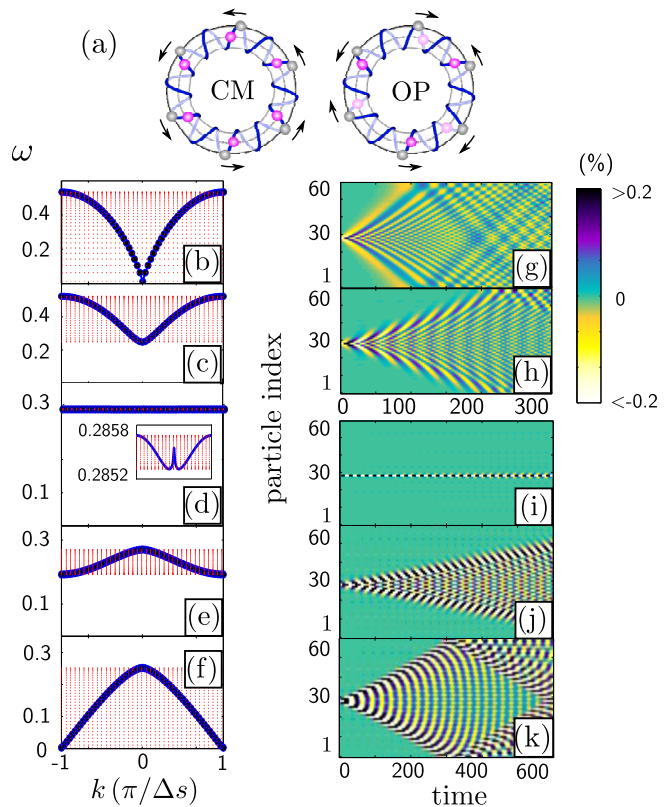}
\end{center}
\caption{\label{disp2} (color online). (a) Sketch of the center of mass (CM, $k=0$) and the out of phase (OP, $k=\pm \frac{\pi}{\Delta s}$) modes.
(b)-(f) Dispersion relation curves $\omega(k)$ for $N=60$ and increasing $r$ corresponding to the points ($1, 2, r_\text{d}, 3, r_\text{cr}$) marked in Fig. \ref{tore1} (c).
For the same values of $r$, panels (g)-(k) present the time evolution following an initial displacement of the particle at $j_0=29$ by $1 \%$ of the equilibrium distance $\Delta s$. 
Colors encode the displacement from equilibrium in units of $\Delta s$. } 

\end{figure}

Results for $\omega(k)$ for different values of $r$ are shown in Fig. \ref{disp2}. 
For  $r=0$ (Fig. \ref{disp2} (b)) 
the  long wavelength  limit $k\rightarrow 0$, corresponding
to identical  displacements of  all particles (CM mode (Fig. \ref{disp2} (a)) has a vanishing frequency $\omega \rightarrow 0$ which follows a linear law
$\omega =v_s k$ (with $v_s$  the sound velocity), resulting from the decoupling of the CM  from the relative motion for the ring geometry.

As the helix radius $r$ increases, the spectrum at small $k$ becomes smoother leading to deviations from the linear
expression and a gap opens at $k=0$  (Fig. \ref{disp2} (c)) due to the coupling of the CM
to the relative motion for $r>0$.
This gap increases with increasing $r$, while the overall width of the spectrum decreases.
At a critical point $r_\text{d}$ (Fig. \ref{disp2} (d)) the spectrum is essentially flat as we have already seen in Fig. \ref{tore1} (c).
A zoom at this point (Fig. \ref{disp2} (d) (inset)) reveals that the degeneracy is very close to, but not complete.
To locate the near-degeneracy point $r_\text{d}$ analytically, we go back to Eq. (\ref{di1}). 
Complete degeneracy would imply that all off-diagonal elements of the Hessian $H_{i,j}, ~ i \neq j$, vanish (the diagonal elements are always identical by symmetry).
Focusing on the nearest-neighbor contributions, we thus find an approximate analytical expression for $r_\text{d}$ by demanding $H_{j,j+1}|_{r_\text{d}}=0$, which yields

\begin{equation}
 r_{_d}=aR\frac{\sqrt{3+\cos(2 a n\pi)}}{\sqrt{2}-a\sqrt{3+\cos(2 a n\pi)}}, \label{codeg2} 
\end{equation}
in excellent agreement with the numerical findings. Indeed, in the thermodynamic limit $R\rightarrow \infty, a\rightarrow 0, aR=\frac{1}{4}$, $r_\text{d}$ tends to $r_\infty=\frac{1}{2\sqrt{2}}$. %as noted before.

Beyond the crossing point, for $r>r_\text{d}$, the curvature of the band changes sign permanently (Fig. \ref{disp2} (e)), 
implying  that the OP mode (Fig. \ref{disp2} (a)) is now lower in frequency than the CM mode.
The width of the spectrum increases again with increasing $r$ until at $r=r_\text{cr}$ the frequency of the OP mode at $k=\pm \pi/ \Delta s$ reaches zero (Fig. \ref{disp2} (f))
and crosses to the imaginary axis for $r > r_\text{cr}$, indicating the symmetric configuration becoming unstable due to the pitchfork bifurcation 
shown in Fig. \ref{tore1} (b).
The condition $\omega(k=\pm \pi/\Delta s)|_{r_\text{cr}} = 0$ can also be tackled analytically, giving an expression for $r_\text{cr}$ which shows that it indeed tends 
to $r_\infty$ in the thermodynamic limit.

The almost full degeneracy of the linearization spectrum at $r_\text{d}$
implies a remarkable localization property in the small amplitude dynamics, 
illustrated in Fig. \ref{disp2} (g) - (k).
Here we explore the time evolution following a 1\% displacement of a single particle at site $j_0$.
Generically, this initially localized excitation spreads over the whole crystal, see e.g. (g) for the case of a ring. 
More precisely, a cone structure emerges indicating a finite velocity at which the excitation proliferates into the crystal. This cone becomes narrower with decreasing bandwidth of the spectrum, see (h), until at the point of near-degeneracy and thus near-zero bandwidth (i) the cone closes and the excitation no longer significantly spreads. 
This unique dynamical feature indicates the presence of an effective screening of interactions at $r=r_\text{d}$, enabling essentially independent motion of the charged particles. 
We emphasize that for this geometric configuration \emph{any} small initial excitation would maintain its shape for large times.
Moving to larger radii $r>r_\text{d}$, the degeneracy is lifted and the bandwidth of the spectrum increases again, thus reopening the cone (Fig. \ref{disp2} (j, k)).\\
Within the linearized equations, the initial dynamics of the spreading can be linked to $\omega (k)$ also on a formal level.
The proliferation of the localized excitation can be quantified by the variance $S(t) =\sum_j j^2 e_j(t) - j_0^2$,
where we employ the local energy $e_j(t)$ at site $j$ as introduced in \cite{Allen1998}, with the time-independent normalization $\sum_j e_j =1$.
Then similar arguments as in \cite{Martinez2012} apply, leading to $S(t) \propto t^2 \int \text d k |\frac{\text d \omega}{\text d k}|^2$,
assuming the crystal is large enough to approximate a sum over $k$ with an integral over the first Brillouin zone.
Consequently, the spreading of an initially localized excitation is ballistic, with a velocity determined
by the square of the group velocity integrated over all $k$. If $\frac{\text{d} \omega}{\text d k}$ is close to zero globally, i.e. the dispersion is almost flat, $S(t)$ will grow only slowly with time and the excitation
will spread only on very long time scales, which is what happens at  $r_\text{d}$.

\begin{figure}[htbp]
\begin{center}
\includegraphics[width=8.6cm]{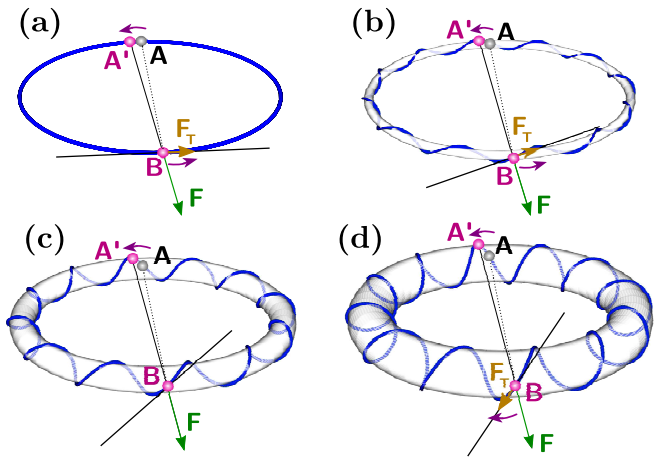}
\end{center}
\caption{\label{forcr} (color online). Schematic illustration of the response of a two particle system $A,B$ to 
a single particle displacement $AA'$ for the cases: (a) $r=0$, (b) $r<r_\text{d}$, (c) $r=r_\text{d}$ and (d) $r>r_\text{d}$. The total force $F$ acting on particle $B$
and its component $F_T$ tangential to the toroidal helix are shown, whereas the curved arrows indicate the directions of displacements.}
\end{figure}

\textit{Degeneracy point} We now provide a geometrical interpretation for the emergence of the degeneracy point $r_d$ in the spectrum.
To this end, let us examine the response of the simplest system of $N=2$ particles, at equilibrium,
confined on the toroidal helix to a single particle displacement (Fig. \ref{forcr}).
A slight counter-clockwise  displacement of particle $A$ towards the position $A'$ results in a force acting on particle $B$.
For the cases $r=0,~ r<r_\text{d}$ (Fig. \ref{forcr} (a),(b)), this force possesses a component tangential to the confining manifold, 
causing a counter-clockwise acceleration of particle $B$.
At  $r_\text{d}$ (Fig. \ref{forcr} (c)), the geometry is such that the displacement $AA'$ results in a force that has no component 
tangential to the toroidal helix curve at the equilibrium position of $B$ and is therefore
entirely compensated by the constraint. Thus, the small amplitude motion of particle $B$ is effectively decoupled from that of $A$.
This simple geometric condition indeed leads to the same value of $r_\text{d}$ as Eq. (\ref{codeg2}) for $N=2$ .
For $r>r_\text{d}$ the force acting on $B$ again attains a non-vanishing projection onto the tangential, but now oriented in the opposite direction, 
causing a clockwise acceleration of particle $B$, in line with the observed inversion of the dispersion relations in this regime (Figs. \ref{disp2} (e),(f)).
For $N>2$, the  geometry parameters can no longer be chosen such that all forces acting on the other particles after displacing a particular one are strictly compensated by the constraint.
Still, it can be seen that the tangential projection of the force acting on particle $j$ after particle $i$ has been slightly
displaced is proportional to the Hessian matrix element $H_{i,j}$. We have seen above that at $r_\text{d}$ all these $H_{i,j} \approx 0$ for $i \neq j$.
Thus, at the point of degeneracy, the geometry 
is such that if a single particle is displaced, 
all resulting force projections onto the local tangents
where the other particles sit are very small simultaneously, resulting in the effective decoupling.

\textit{Conclusions} We have shown that for charged particles confined 
on a 1D toroidal helix, a linear to zig-zag-like bifurcation occurs when increasing 
the radius of the helix at commensurate fillings. Such kinds of bifurcations are typical of Wigner crystals under harmonic trapping 
\cite{Fishman2008,Cartarius2013} and are attributed to the increment of dimensionality from 1D to 3D. In our case, however, the single particle configuration space remains always strictly 1D,
a fact that manifests itself in the way the critical value $r_\text{cr}$ is reached. 
In particular, for the transition to occur, the OP mode resulting in zig-zag deformations of the crystal has to cross zero at $r_\text{cr}$,
in contrast to the ring limit $r=0$ where the OP mode has the largest frequency. 
This necessarily implies an inversion of the dispersion relation's curvature when approaching $r_\text{cr}$, 
as the reduced dimensionality precludes a transverse branch which usually causes the bifurcation \cite{Fishman2008}.
Notably, the deformation of the dispersion curve when increasing $r$ towards $r_\text{cr}$ passes through a point where all modes are essentially degenerate and the dispersion is flat.
For this particular geometry of the constraint manifold, the (small amplitude) dynamics of the particles is effectively decoupled,
allowing for localized excitations that do not spread into the crystal.
Such intriguing properties may be of interest in phononics, regarding the properties of acoustic meta-materials \cite{Deymier2013} 
and applications such as sound isolation and cloaking \cite{Maldovan2013} or
even information storage  \cite{Apollaro2007}.
Beyond these, the present setup offers multiple possibilities to control the vibrational band structure,
rendering it an attractive device e.g. in the context of free-standing helical nanostructures.

\begin{center}
{ \textbf{ACKNOWLEDGEMENTS}}
\end{center}
A. Z. thanks C. Morfonios and P. G. Kevrekidis for fruitful discussions
and the International Max Planck Research School for Ultrafast Imaging and Structural Dynamics for a PhD scholarship.
J. S. gratefully acknowledges a scholarship from the {\it Studienstiftung des deutschen Volkes}.

\end{document}